# The Equality Maturity Model: an actionable tool to advance gender balance in leadership and participation roles

Paloma Díaz
Department of Computer Science and Engineering
Universidad Carlos III
Leganés, Spain
pdp@inf.uc3m.es
0000-0002-9493-7739

Paula Alexandra Silva
DEI/CISUC
Universidade de Coimbra
Coimbra, Portugal
paulasilva@dei.uc.pt
0000-0003-1573-7446

Katja Tuma
Department of Computer Science
Vrije Universiteit Amsterdam
Amsterdam, Netherlands
k.tuma@vu.nl
0000-0001-7189-2817

*Abstract*— The underrepresentation of women in Computer Science and Engineering is a pervasive issue, impacting the enrolment and graduation rates of female students as well as the presence of women in leadership positions in academia and industry. The European Network For Gender Balance in Informatics (EUGAIN) COST action seeks to share data, experiences, best practices, and lessons from failures, and to provide actionable tools that may contribute to the advancement of gender balance in the field. This paper summarises results from the Ph.D./Postdoc to Professor workgroup that were gathered in two booklets of best practices. Specifically, we introduce the Equality Maturity Model (EMM), a conceptual tool aimed at supporting organisations in measuring how they are doing concerning equality and identifying potential areas of improvement and that was inspired by both booklets.

I. INTRODUCTION

The lack of women in Computer Science and Engineering is a pervasive issue, impacting the enrolment and graduation rates of female students as well as the presence of women in leadership positions both in academia and industry. As the She Figures 2021 report shows, women are underrepresented in STEM fields, and this situation is even more severe in the Information and Communication (ICT) disciplines with only 20.4% of women graduating with an ICT bachelor [1]. The gap between women and men is also evident in research and leadership positions [1].

Cultural and organizational biases, the so-called *second-generation discrimination*, as well as problems like the leaking pipeline, the glass ceiling, the sticky floor, and the glass cliff, among others, have been extensively studied in the literature [2, 3]. The creation of Diversity, Inclusion, and Equity (DEI) committees and gender equality plans is being pushed everywhere, from institutions to professional organisations such as IEEE or ACM. The European Union even made Gender Equality plans mandatory for institutions participating in H2020 calls. However, dealing with gender underrepresentation may not be only a question of generating documents or creating ad-hoc committees, nor of changing women by applying the Henry Higgins effect [4], but of promoting changes in the organisational cultures and society to make people aware that diversity and gender equality are valuable assets in a sustainable and fair society.

As stated in [5], without specific interventions the lack of equality between men and women, not only in terms of representation but also of leadership and salary, will not be solved in this century.

The European Network For Gender Balance in Informatics (EUGAIN) COST action[1] is a step forward in this direction as it aims at providing data, experience on best practices and failures, and actionable tools for advancing gender balance in computer science. In particular, workgroup 3 (WG3) is devoted to dealing with the path from the Ph.D./Postdoc to Professor, trying to provide useful resources to increase the leadership and participation of women in academia and also in professorship and leadership positions. In this paper, we introduce one of the actionable tools devised in this workgroup, the Equality Maturity Model (EMM, henceforth), a conceptual tool aimed at helping organisations measure how they are currently doing concerning equality and identifying potential areas of improvement.

In the next section, we review the related work that inspired the EMM. Section III provides context for this work describing the main goals of the EUGAIN COST Network and, more specifically, the outcomes of WG3 - From Ph.D./Postdoc to Professor. Then we describe the current version of EMM, which is still under development, and, finally, we draw initial conclusions and future directions to improve the model and assess it.

II. . RELATED WORK

The gender gap in the workforce and leadership positions has already been identified as a problem in many academic and research institutions not only because it suggests a lack of equity in hiring and promoting women, but also because diversity increases creativity, profitability, and performance [6, 7]. However, these benefits strongly depend on the existence of a supportive organizational culture [8] aimed at valuing diversity and not accepting it as a buzzword or a legal imposition. Organizational change requires commitment from the governing boards as well as acknowledging whether and how the current situation can be changed.

The EMM aims to provide an actionable tool to think about the gender equality policies, strategies, and actions that are currently implemented and the improvements that could be applied to make progress through the EMM. Inspired by the Capability Maturity Model Integration (CMMI®) that is applied in software engineering [8], the EMM is a conceptual tool designed to assist organisations in evaluating their current status regarding gender equality and to help identify areas where improvements are possible.

---

[1] eugain.eu

The CMMI® [8] was created by the Software Engineering Institute at Carnegie Mellon University to provide organisations with a set of best practices to improve their processes (including requirements analysis, software development, systems engineering, project management, etc.), and the quality of the products and services they offer. Both aspects make up the concept of *maturity*. CMMI® is a framework that offers tools to assess and improve software development maturity in a systematic way. To start with, it identifies five increasing levels of maturity that are applied to the various processes carried out in the organisation:

1. *Initial*: Processes are unpredictable, chaotic, and rarely controlled to the point that any success essentially depends on individual efforts and intuitions.

2. *Managed*: Most processes are defined using projects or tasks that respond to very specific goals aimed at coping with already identified issues. The organization has mainly a reactive behaviour though there are controls at some points in the process.

3. *Defined*: Processes are well characterised and understood by all the implied stakeholders. Processes are proactive and tailored to the organization's needs and are described in a detailed way using standards, procedures, tools, and methods. At this level, there is a focus on continuously learning and improving the process.

4. *Quantitatively Managed*: Processes are measured and controlled with objective and meaningful data. Quantitative data make it possible to see and manage performance and quality in terms of statistical measures. This approach is applied during the whole life cycle of the processes.

5. *Optimizing*: Processes focus on continuous improvement based on a quantitative understanding of the common causes of variation and the impact of improvement opportunities. The organization is proactive in identifying opportunities for improvement and it is agile in implementing such improvements.

Apart from the levels, the CMMI® framework offers a conceptual scaffold to help organisations apply this model to different processes related to software development and management. The ultimate goal is to systematize the analysis of the current situation and the identification of areas of opportunity to improve current practices.

This same approach of systematization and process maturity has been applied in other domains. Thus, the People Capability Maturity Model [9] is aimed at helping human resources departments in setting "*basic workforce practices that can be continuously improved to develop the capability of the workforce*" [9]. In this case, best practices and measures are identified for related processes such as staffing, communication and coordination, work environment, performance management, training and development, and compensation. Similarly, the Digital Transformation Maturity Model [10] identifies several processes and dimensions to systematically analyse this process.

This level of structured systematization in the process evaluation might be what is required in institutions and organisations regarding gender equality. This is the purpose of providing institutions and organisations with a tool with which to comprehensively analyse their current actions and, what is even more important, the actions they are not yet doing and that could have a great impact in terms of gender equality. In this way, the EMM could be a propeller for action towards gender balance.

### III. THE EUGAIN COST NETWORK

The EUGAIN COST action aims to provide data, experiences, and tools for advancing gender balance in Informatics. EUGAIN is organised into five workgroups that reflect the different stages where specific actions can be taken, including:

- WG1. From School to University, that is focused on how to increase the number of female students in Informatics and, in general, Information and Communication Technologies (ICT) bachelors and how to support them to actively participate and succeed in their path.

- WG2. From Bachelor/Master studies to PhD, which is centred on the next stage where the gap is increased, the number of PhD in ICT.

- WG3. From Ph.D./Postdoc to Professor, which focuses on specific problems and challenges that make the academic career of women more difficult and how to address these challenges from the institutions.

- WG4. The cooperation with industry and society aims at integrating all the initiatives regarding gender balance in which industry and relevant stakeholders are being involved.

- WG5. Strategy and Dissemination, whose goal is to raise awareness and advocate for change, reach out to external stakeholders, and disseminate the results of the Action.

The EMM model introduced in this paper is a result of WG3's collaborative efforts. It builds on prior work of the team members, specifically two booklets: i) a guide on best practices for aiding the transition of PhD and postdoctoral researchers into faculty positions [12], and ii) a booklet focused on career planning and mentoring to enhance organizational support initiatives [13].

The first booklet is aimed at university departments and highlights gender balance issues, addressing the reasons for women's underrepresentation and proposing strategies for improvement in recruiting, retaining, and promoting women, along with guidelines for evaluating hiring and promotion applications. These practices are intended for application throughout an academic career, promoting equal opportunities for women.

The second booklet is aimed both at management and women and examines structural factors, discussing the prevalent culture in academic institutions, to then suggest specific actions to address these factors. Steps into action are presented to foster a supportive environment, implement career development initiatives, and establish mentoring programs. The EMM emerged as a mechanism to systematize the analysis of the current status of institutions and organisations and to align the application of the identified strategies and practices with strategic goals and processes.

## IV. THE EQUALITY MATURITY MODEL

The EMM aims to become an actionable tool for institutions and organisations to reflect upon the actions they are taking, that they could take in the future to promote gender equality and the mechanisms they could implement to support the required cultural change, awareness, and accountability.

Similarly to CMMI®, the EMM aims at providing institutions and organisations with a snapshot of their maturity regarding equality policies, strategies, and practices as well as hints on how to progress to higher levels. The EMM includes Policies, Strategies, and Practices (see Figure 1). *Policies* are principles assumed in the organisation, *strategies* are action plans to achieve policies, and *practices* are specific and well-organised activities that make up the strategies. All of them need to be clear, known by the community, accountable, and open to improvement.

For instance, a broadly stated *policy* is Promoting Gender Equality, which is also the United Nations' Sustainable Development Goal 5: "Achieve gender equality and empower all women and girls". There are many ways in which an organisation can achieve this objective, depending on the specific goals the organisation aims to reach, the available resources, existing barriers and opportunities inside and outside the organisation, and the starting/current point at which the organisation finds itself in terms of gender equality. The point at which the organisation finds itself is what we identify with the concept of maturity of the organisation in gender equality. From such a general goal, a potential *strategy* could be to increase the number of women professors in STEM-related departments. In turn, several *practices* can be implemented for such a strategy, including promoting gender-inclusive hiring guidelines, creating inclusive working environments, adopting positive discrimination, using promotion criteria inclusive of different career paths and that value different types of contributions, and so on.

By prompting the organisation to position itself and to set up and reflect upon policies, strategies, and practices, the EMM can help to diagnose where the institution is concerning a policy, which strategies should be prioritised, and which action can be put into practice.

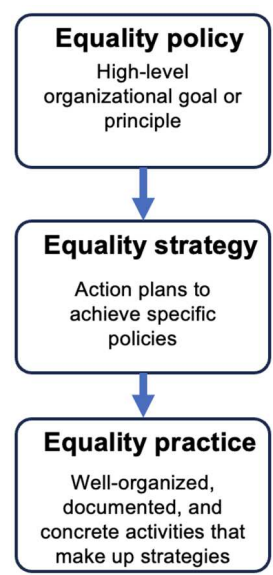

Fig. 1. Levels of description at the EMM

The EMM is structured in five relevant dimensions that permeate different processes at the institution. To identify the current dimensions, we complement the outcomes of the two previous deliverables of WG3 [12, 13], with a literature review of existing problems and practices [2, 6, 14-18]. The dimensions currently included in EMM are the following:

- *Organisation Policy on Equality*: This dimension focuses on the specific actions carried out in the organization to support the required culture change and the right conditions to turn gender equality into a valuable asset. Cultural change needs to be acknowledged and valued by all to avoid negative effects [13]. This implies raising awareness of the value of gender equality among all the involved agents in a continuous way. A one-shot training program might not be enough as the culture of equality has to reach everyone in the organisation during its whole lifecycle. It also needs to go deeper than just raising awareness as it needs to change the organisation's values and beliefs. For example, the culture of science that could be perceived as neutral, since it relies upon "objective" data, does not take into account that by rewarding only individual and competitive merits, the contribution of multidisciplinary teamwork is not acknowledged, even if it nurtures innovation [2]. Another aspect to analyse is whether the organisations have a gender-balanced policy in leadership and decision-making that does not overwhelm women with useless work and does not push them to the glass cliff [2].

- *Inclusive Work Environment*: One of the main causes of the lack of women in leadership positions is attributed to what is called the "chilly climate" [14] for women in academia and research [2, 6]. Among other issues, this dimension focuses on the recognition of different kinds of contributions and career paths, actions for work-life balance (including flexible work arrangements or disconnection policies), standards of conduct to avoid gender-based violence or discrimination, and identification of gendered experiences that challenge preconceived beliefs about the organisation. For instance, the qualitative study carried out in the NIST [2] showed that the organizational values of excellence and meritocracy contributed to nurturing an endemic elitism that not only favoured men but also was not necessarily aligned with the contribution to the goals of the institution. An inclusive working environment starts by making the job attractive to women and other unrepresented groups when advertising new openings or promotions [12].

- *Fair Recruitment, Promotion, and Evaluation Processes*: It has been extensively documented that women experience several micro-inequalities in academia and research that harden their careers. Micro-inequalities were identified back in the 90s to refer to all the small "*things*" that happen covertly and that correspond to biases against a minority, whether conscious or not [15]. Women are usually perceived as not as good scientists as men [2], have fewer opportunities to participate in panels and keynotes, are less credited in scientific papers and patents [16], and their work receives fewer citations [2]. Women are not less productive than men, but they value and engage in other kinds of activities, often voluntary tasks such as mentoring, that are as important and needed to reach the goals of the organizations. Indeed, it is also recognised that women volunteer and are directly assigned more non-promotable tasks [9] losing part of their working time not doing the research that is really valued in the committees. Hence evaluation criteria must recognize different types of contributions and not only numbers as well as different career paths [2, 12]. There are also differences in the recommendation letters written for men and women [17] and hiring and promotion committees might have unconscious biases that favour men. Many recommendations can be applied when hiring or promoting, as described in [12].

- *Career Development Program*: One of the gendered experiences that was identified in [2] was a different perception of how to progress in the professional career that seemed quite dependent on gender, mainly due to the chilly climate and the hidden micro-inequalities. Implementing specific career development plans can help minimize the effect of some of the biases described in the previous dimensions by providing specific networking and visibility opportunities for women's contributions [2,12] whilst it will help in having a clear, equal and transparent career description within the organization [12]. Mentoring programs are also a key instrument to support women in their career development [13].

- *Gender Equality Plans*: Gender equality plans are almost mandatory in most European Universities, especially after the European Union obliged all institutions to have a gender equality plan to participate in H2020 calls. This dimension not only focuses on the binary fact of having or not having an equality plan but on how mature it is, how it is implemented, how it permeates different areas in the organization, and how this impact is measured [18].

These above-described dimensions are, like the CMMI® model, assessed using five levels. Each level is characterized by a rubric so that institutions can reflect upon where they currently stand and how they could go forward if needed. Thus, examples of practices at each level will be also provided to assist institutions in identifying ways to progress in their policies, strategies, and practices. As an example, Figure 1 shows a simplified rubric for the Gender Equality Plan dimension that is explained below.

**Does your institution/organization have an equality plan?**

| Level | Description |
|---|---|
| Initial | We don't have an Equality Plan or we have a basic equality plan not publicly available or not known by all the organization |
| Managed | We have a detailed equality plan that is publicly available, but there are no specific actions to raise awareness on it |
| Defined | We have a detailed and publicly available equality plan and we carry out specific actions to raise awareness in all the community, promote participation of different stakeholders and control its application in different processes |
| Quantitavely Managed | We have a detailed equality plan, we carry out actions to raise awareness on its application for all the community and we check for compliance in the involved processes getting quantitative data about the impact of the measures included in the plan |
| Optimizing | We have a detailed equality plan, we carry out actions to raise awareness on its application for all the community, we check for compliance in the involved processes and there is a specific committee that evaluates its applicability, quality and identifies areas of improvement |

Fig. 2. Example of a rubric in the EMM

Having an equality plan serves as a mere checkpoint that does not demonstrate the institution's maturity concerning equality strategies and practices. At the *Initial level*, a plan may be in place, possibly due to mandatory requirements, but the entire organisation lacks awareness of the plan and does not align its work and behaviours to the plan. Specific strategies or practices to implement are not identified at this level, making it impossible to ensure whether the plan has any impact on gender equality.

At the *Managed Level*, the plan is outlined in terms of generic strategies with no clear ways of acting to propel progress. While the plan is known to all within the organisation, it may not be because is published but rather because due to an informal learning process that has taken place. Furthermore, modifications to the plan only occur when faced with 'awkward' situations, prompting the institution to act upon it and respond.

Moving to the *Defined level*, the plan includes well-defined strategies and potential practices or actions. The plan is acknowledged and agreed upon by all involved stakeholders, reflecting the cultural shift the institution is embarking on. The plan is not solely derived from internal issues but also from proactively learned lessons from other institutions, to anticipate problems and define how to prevent them and address them. Continuous learning and training play a pivotal role in this stage to enhance the plan's comprehensiveness.

At the *Quantitatively managed level*, the institution has identified specific criteria, metrics, or success indicators to objectively measure the actual impact of the applied strategies and practices. The management team can demonstrate that progress is a result of a specific policy based on objective measures, establishing a clear link between data and actions. Using data with no direct relation to a policy might imply a casual situation rather than a causal consequence.

Finally, at the *Optimizing level*, the institution continuously and quantitively monitors its managed level plan, learning about successful and failed practices. At this level, the aim is to be proactive and agile in improving the plan, demonstrating a commitment to ongoing refinement of the plan.

In this short paper, we propose the EMM as a tool to facilitate cultural change. For such a ground-breaking shift to materialize in organizations and institutions, the political, social, and ideological readiness of society is still necessary. Future work will focus on evaluating the effectiveness of the EMM in practice and on gaining insight into the best way to apply it in real-life scenarios. To investigate these aspects is planned for future work. The recommended way to use the EMM is to establish exploratory focus groups involving different stakeholder profiles in terms of gender and responsibilities to discuss, openly and respectfully, the status of the institution regarding the different dimensions. Having a varied set of participants will enrich the process, as the perception of different stakeholders will provide a better picture of the current situation and how it is experienced by diverse people. Though managers could think that their institution is fully committed to gender equality, hearing the voices of those who daily experience the working environment can provide a completely different perspective, as shown in the study carried out by Theofanos *et al.* [2].

## V. Conclusions and Future Work

A cultural change is needed to close the gender gap in academia and research. This change has to involve the whole organization and that must be perceived as a joint commitment to advance as an institution, not just as a legal or social imposition. As part of the EUGAIN network, we described here some of the outcomes of WG3, when studying possible ways forward in the career path from Ph.D./Postdoc to Professor. In particular, we introduced a conceptual tool that is currently being developed to assist organisations in engaging in a self-reflection and improvement process, the EMM. Though the EMM is in a preliminary stage and is still open to growth with best practices and strategies, it represents an important first stage to provide a systematic and actionable tool to reflect upon the maturity of institutions in terms of gender equality. While we recognise that the EMM tool will need yet to be evaluated by institutions and organizations to verify its effectiveness, this paper marks an important initial effort to introduce a more systematic measurement of gender equality. The model, based on discussions in heterogeneous focus groups, will also contribute to raising awareness in the whole organisation about the importance of gender equality and identifying the hidden biases and micro-inequalities in the institution.

ACKNOWLEDGEMENT

This work has been partially supported by the EUGAIN COST Action CA19122— European Network for Gender Balance in Informatics.